\documentstyle[prb,aps,twocolumn,psfig]{revtex}

\def\bS{{\mbox{\boldmath $S$}}}
\def\bi{{\bf i}}
\def\bj{{\bf j}}
\def\bd{{\bf d}}

\begin{document}
\wideabs{
\title{Heisenberg antiferromagnet on the square lattice for $S\ge{1}$}

\author{Alessandro Cuccoli, Valerio Tognetti and Paola Verrucchi}
\address{Dipartimento di Fisica dell'Universit\`a di Firenze
	and Istituto Nazionale di Fisica della Materia (INFM),
	\\ Largo E. Fermi~2, I-50125 Firenze, Italy}

\author{Ruggero Vaia}
\address{Istituto di Elettronica Quantistica
	del Consiglio Nazionale delle Ricerche,
	via Panciatichi~56/30, I-50127 Firenze, Italy
	\\and Istituto Nazionale di Fisica della Materia (INFM),
	Largo E. Fermi~2, I-50125 Firenze, Italy}

\date{\today}

\maketitle

\begin{abstract}
Theoretical predictions of a semiclassical method -- the pure-quantum
self-consistent harmonic approximation -- for the correlation length and
staggered susceptibility of the Heisenberg antiferromagnet on the square
lattice (2DQHAF) agree very well with recent quantum Monte Carlo data for
$S=1$, as well as with experimental data for the $S=5/2$ compounds
Rb$_2$MnF$_4$ and KFeF$_4$.
The theory is parameter-free and can be used to estimate the
exchange coupling: for KFeF$_4$ we find $J=2.33\pm0.33$~meV,
matching with previous determinations.
On this basis, the adequacy of the quantum nonlinear $\sigma$ model
approach in describing the 2DQHAF when $S\ge{1}$ is discussed.
\end{abstract}
}

We consider the Hamiltonian
\begin{equation}
 \hat{\cal{H}}= {J\over2} \sum_{{\bf{i}},{\bf{d}}}
 \hat {\mbox{\boldmath$S$}}_{\bf{i}} \cdot
 \hat {\mbox{\boldmath$S$}}_{{\bf{i}}+{\bf{d}}}~,
\label{e.H}
\end{equation}
where $\bi\equiv(i_1,i_2)$ runs over the sites of a 
square lattice, $\bi+\bd$ defines the four nearest neighbours 
$(i_1\pm 1,i_2\pm 1)$ and the quantum operators $\hat{\bS}_\bi$ obey the 
angular momentum commutation relations 
$[S_\bi^\alpha,S_\bj^\beta]=iS_\bi^\gamma\delta_{\bi\bj}
\varepsilon^{\alpha\beta\gamma}$ with $|\hat{\bS}_\bi|^2=S(S+1)$.
When $J>0$ Eq.~(\ref{e.H}) describes the 2D quantum Heisenberg 
antiferromagnet (QHAF).

From the experimental point of view the interest in this model is due to 
the existence of several compounds that, despite their complex structure,
can be described by the Hamiltonian (\ref{e.H}) as far as their 
magnetic behaviour is concerned\cite{SokolP93}. 
Some of them are parent compounds of high-$T_{\rm c}$ superconductors,
which makes the study of their thermodynamic properties of particular 
relevance. Another interesting feature of this class of materials is that
it contains compounds with $S=1/2$ (La$_2$CuO$_4$, Sr$_2$CuO$_2$Cl$_2$),
$S=1$ (La$_2$NiO$_4$, K$_2$NiF$_4$) and $S=5/2$ (KFeF$_4$, Rb$_2$MnF$_4$) 
thus allowing an experimental analysis of the 2DQHAF as its quanticity varies 
from the extreme quantum case $S=1/2$ to the almost classical one $S=5/2$.
This is of particular interest when a comparison with theoretical results is 
attempted, as the spin value $S$ appears as a parameter that can be easily
changed in most theoretical approaches, obviously save the numerical 
simulations.

The theory of the QHAF has been generally related with that of the 
quantum nonlinear $\sigma$ model (QNL$\sigma$M) by
Chakravarty, Halperin and Nelson (CHN)\cite{ChakravartyHN89}
whose work led to the first direct comparison between experimental 
data and the results of the QNL$\sigma$M field theory; the surprisingly 
good agreement found caused an intense activity, both theoretical and
experimental, about the subject.
In the last few years, however, it turned out that for larger spin
neither the CHN formulas nor the improved ones derived by Hasenfratz and
Niedermeier (HN)\cite{HasenfratzN91}, 
can  reproduce the experimental data.
The discrepancies observed may be due to the fact
that the real compounds do not behave like 2DQHAF or to an actual 
inadequacy of the theory. In particular the CHN-HN scheme introduces
two possible reasons for such inadequacy to occur: the physics of the
2DQHAF is not properly described by that of the 2DQNL$\sigma$M and/or
the two(three)-loop renormalization-group expressions derived by CHN(HN)
do hold at temperatures lower than those experimentally accessible.

The situation can be clarified by using an independent 
theoretical method, directly applicable to the QHAF and whose validity in 
the temperature region of interest can be checked.
The high-temperature expansion (HTE) technique is characterized by such  
requisites and the first well sound doubts about the QNL$\sigma$M
picture of the QHAF indeed arose from HTE results 
\cite{Elstner97Etal95};
however, the HTE is not applicable in the whole temperature range where 
data from experiments or QMC simulations are available, so that it 
cannot be used to develop a complete analysis of the subject. 
On the other hand, such analysis can be carried out by means of the 
pure-quantum self-consistent harmonic approximation 
(PQSCHA)\cite{CGTVV95}, whose results for $S\ge 1$ are fully reliable at 
all temperatures (except the extremely low ones in the $S=1$ case), 
as we show below.

The PQSCHA is based on the path-integral formulation 
of quantum statistical mechanics, and has been successfully applied to many
magnetic systems\cite{CGTVV95}; 
it permits to express quantum thermal averages 
of physical observables in the classical-like form of phase-space integrals,
where the integrand functions, depending on both $T$ 
and $S$, are determined 
from the quantum operators according to  a precise procedure.
The final formulas for the thermodynamic quantities do not contain 
parameters other than those appearing in the Hamiltonian of the model under 
investigation. In the case of magnetic systems the method applies 
directly to the spin model, with the spin value $S$ appearing as a 
coupling parameter that can be easily varied. The PQSCHA expression for the
statistical average of a physical observable described by the quantum
operator $\hat O$ is given by the phase-space integral
$
\langle\hat {\cal O}\rangle=(1/{\cal Z})
 \int d^{\scriptscriptstyle N}\!{\mbox{\boldmath$s$}}
 ~\widetilde{\cal O}~
 \exp(-\beta {\cal H}_{\rm{eff}})
$
where $\beta=T^{-1}$;
$N$ is the number of lattice sites,
${\mbox{\boldmath$s$}}$ is a classical
vector on the unitary sphere  ($|{\mbox{\boldmath$s$}}|=1$) and
${\cal{Z}}{=}\int d^{\scriptscriptstyle N}\!{\mbox{\boldmath$s$}}
 ~\exp(-\beta{\cal{H}}_{\rm{eff}})$ is the partition function. 
The determination of
the effective Hamiltonian 
${\cal{H}}_{\rm eff}=
{\cal{H}}_{\rm eff}(\{{\mbox{\boldmath$s$}}_{\bf{i}}\})$
represents the core of the application of the PQSCHA method.
The function 
$\widetilde{\cal{O}}=\widetilde{\cal{O}}
(\{{\mbox{\boldmath$s$}}_{\bf{i}}\})$
is obtained starting from the quantum operator $\hat O$
and following the same procedure used to determine the configurational 
part of ${\cal{H}}_{\rm eff}$. 

The energy scale $J\widetilde{S}^2$ with 
$\widetilde{S}\equiv S+1/2$ naturally appears
in deriving the effective Hamiltonian, and we hence define, 
and hereafter use, the reduced temperature $t\equiv T/J\widetilde{S}^2$.
In the specific case of the 2DQHAF described by Eq.~(\ref{e.H})
we find
\begin{equation}
 {{\cal H}_{\rm{eff}}\over J\widetilde{S}^2}
 = {\theta^4\over2} \sum_{{\bf{i}},{\bf{d}}} {\mbox{\boldmath$s$}}_{\bf{i}}
 {\cdot} {\mbox{\boldmath$s$}}_{{\bf{i}}+{\bf{d}}} + {\cal G}(t)~,
\label{e.Heff}
\end{equation}
where $\theta^2=1-{\cal D}/2$,
${\cal D}=\sum_{\bf{k}}
 (1-\gamma_{\bf{k}}^2)^{1/2}{\cal L}_{\bf{k}}/(N{\widetilde{S}})$,
${\cal{L}}_{\bf{k}}=\coth f_{\bf{k}}-f_{\bf{k}}^{-1}$,
$f_{\bf{k}}=\omega_{\bf{k}}/(2\widetilde{S}t)$, 
$\gamma_{\bf{k}}=(\cos{k_1}+\cos{k_2})/2$ and 
${\bf{k}}\equiv(k_1,k_2)$ wave vector in the first Brillouin zone;
${\cal{G}}(t)$ is a uniform term that does not affect the evaluation of 
statistical averages.
The self-consistent solution of the two coupled equations
$
\omega_{\bf{k}}=4 \kappa^2 (1-\gamma^2_{\bf k})^{1/2}$ and 
$
\kappa^2=\theta^2-t/(2\kappa^2)
$
gives us all the ingredients needed to evaluate the thermodynamic
properties of the system. Details about the derivation of the 
effective Hamiltonian and of the above formulas are given 
in Ref.~\onlinecite{CTVV96prl}.

The renormalization coefficient ${\cal{D}}={\cal{D}}(S,t)$
 measures the strength of 
the pure-quantum fluctuations, whose contribution to the thermodynamics 
of the system is the only approximated one in the PQSCHA scheme:
The theory is hence quantitatively meaningful as far as 
${\cal{D}}$ is small enough to justify the self-consistent harmonic treatment
of the pure-quantum effects.
In particular the simple criterion ${\cal{D}}< 0.5$ 
is a reasonable one to check the validity of the final results.

\begin{figure}[hbt]
\centerline{\psfig{bbllx=16mm,bblly=69mm,bburx=192mm,bbury=207mm,%
figure=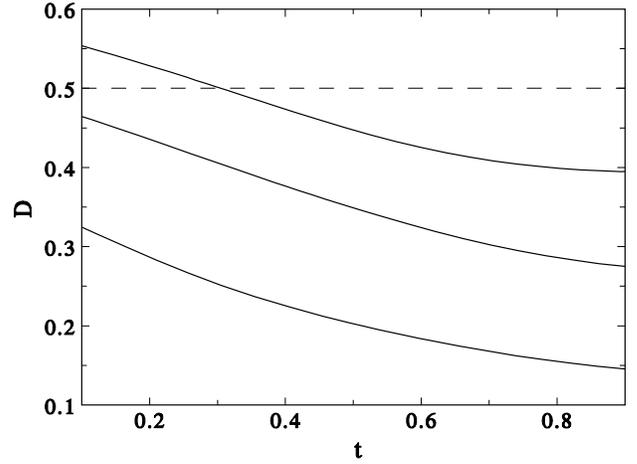,width=82mm,angle=0}}
\caption{Renormalization coefficient ${\cal{D}}$ vs $t$ for  
$S=1/2$, $1$, $5/2$ (from the top curve).
\label{f.D}
}
\end{figure}

In Fig.~\ref{f.D} we show the coefficient ${\cal{D}}(S,t)$ as a function of
temperature and for different spin values: Besides the obvious observation
that the temperature range where ${\cal{D}}<0.5$ depends on the spin value,
we also note that for $S=1$ such interval extends to almost the whole 
temperature range, leaving the extreme quantum case $S=1/2$ the 
only delicate one as far as the validity of the PQSCHA is concerned.
Fig.~\ref{f.D} 
should clarify that the PQSCHA is not a high-temperature method, 
but rather a semiclassical one whose results are fully reliable 
already for $S=1$.

The $S=1/2$ case is extensively discussed in Ref.~\onlinecite{CTVV97prlre} 
and we do not deal with it in this paper; however, we
note that our $S=1/2$ results agree with QMC data for $t\ge 0.4$
which means, as seen in Fig.~\ref{f.D}, ${\cal{D}}\le 0.47$.
This confirms the criterion adopted to be well sound.

From Eq.~(\ref{e.Heff}) 
we see that the symmetry of the Hamiltonian is left unchanged 
so that the quantum system essentially behaves,
at an actual temperature $t$, as its classical counterpart does 
at an effective temperature
$t_{\rm eff}=t/\theta^4(t)~$.
Being $\theta^4(t)<1$, we see that the system is more disordered
than its classical counterpart, as the pure-quantum fluctuations make
the former behave like the latter does at a higher temperature.  

As for the correlation functions 
$G({\bf{r}})\equiv\langle\hat{\mbox{\boldmath$S$}}_{\bf{i}}
{\cdot}\hat{\mbox{\boldmath$S$}}_{{\bf{i}}+{\bf{r}}}\rangle$, with
${\bf{r}}\equiv(r_1,r_2)$ any vector on the square lattice, we find 
$G({\bf{r}})= \widetilde{S}^2 \theta^4_{\bf{r}}
 \langle{\mbox{\boldmath$s$}}_{\bf{i}}
 {\cdot}{\mbox{\boldmath$s$}}_{{\bf{i}}+{\bf{r}}}\rangle_{\rm{eff}}$
where $\langle~\cdot~\rangle_{\rm eff}$ is 
the classical-like statistical average with 
the effective Hamiltonian; the parameter 
$\theta^4_{\bf{r}}=1-{\cal{D}}_{\bf{r}}$, 
defined in Ref.~\onlinecite{CTVV96prl},
goes to a constant and finite value for large $|{\bf{r}}|$.
From the above formulas, the correlation length, defined from
the asymptotic expression $G({\bf{r}})\propto\exp(-|{\bf{r}}|/\xi)$ for
large $|{\bf{r}}|$, turns out to be
\begin{equation}
\xi(t)=\xi_{\rm cl}(t_{\rm eff})
\label{e.xi}
\end{equation}
where $\xi_{\rm cl}$ is the correlation length of the classical HAF,
unique ingredient we need to obtain $\xi(t)$ for the quantum 
system, being the evaluation of $\theta^4(t)$ a simple matter for any
spin value. 
The PQSCHA expression for the staggered 
susceptibility $\chi\equiv\sum_{{\bf{r}}}(-)^{r_1+r_2}~G({\bf{r}})/3$ is
$$
 \chi={1\over 3}\bigg[
 S(S+1) + \widetilde{S}^2 \sum_{{\bf{r}}\neq 0} (-)^{r_1+r_2}
 ~\theta^4_{{\bf{r}}} ~\langle{\mbox{\boldmath$s$}}_{\bf{i}}
 {\cdot}{\mbox{\boldmath$s$}}_{{\bf{i}}+{\bf{r}}}\rangle_{\rm{eff}} \bigg]~,
$$
and in this case we need to know the classical $G({\bf{r}})$ at any
${\bf{r}}$ to obtain the numerical value of the quantum $\chi$.

\begin{figure}[hbt]
\centerline{\psfig{bbllx=16mm,bblly=69mm,bburx=192mm,bbury=207mm,%
figure=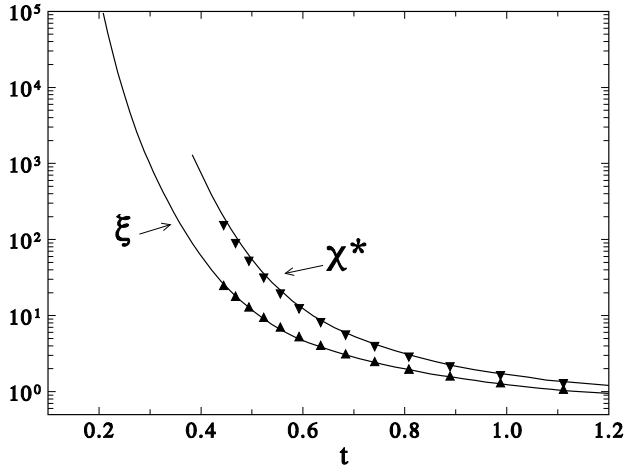,width=82mm,angle=0}}
\caption{Correlation length $\xi$ and staggered 
susceptibility $\chi^*\equiv \chi/\widetilde{S}^2$
 vs $t$ for $S=1$. Symbols are QMC data from
Ref.~\protect\onlinecite{HaradaTK98}.
\label{f.xichis10}
}
\end{figure}

\begin{figure}[hbt]
\centerline{\psfig{bbllx=16mm,bblly=69mm,bburx=192mm,bbury=207mm,%
figure=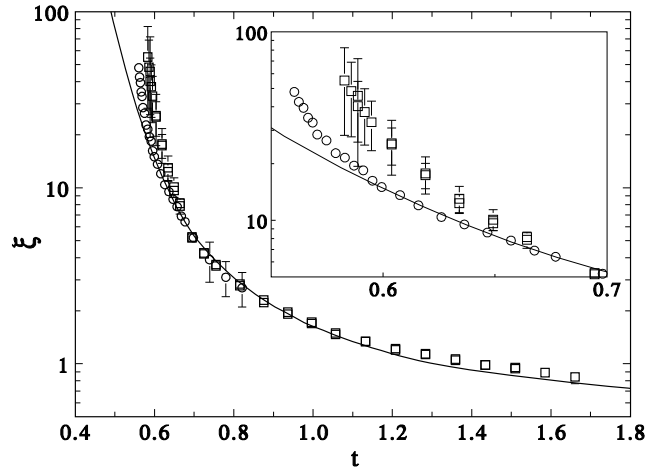,width=82mm,angle=0}}
\caption{Correlation length $\xi$ vs $t$ for $S=5/2$.
Symbols are experimental data 
for KFeF$_4$ (circles)\protect\cite{FultonEtal94}
 and Rb$_2$MnF$_4$ (squares)\protect\cite{LeeEtal98}. The inset shows 
 a magnification of the region where the crossover between Heisenberg and
 Ising behaviour is observed.
\label{f.xis25}
}
\end{figure}

\begin{figure}[hbt]
\centerline{\psfig{bbllx=16mm,bblly=69mm,bburx=192mm,bbury=207mm,%
figure=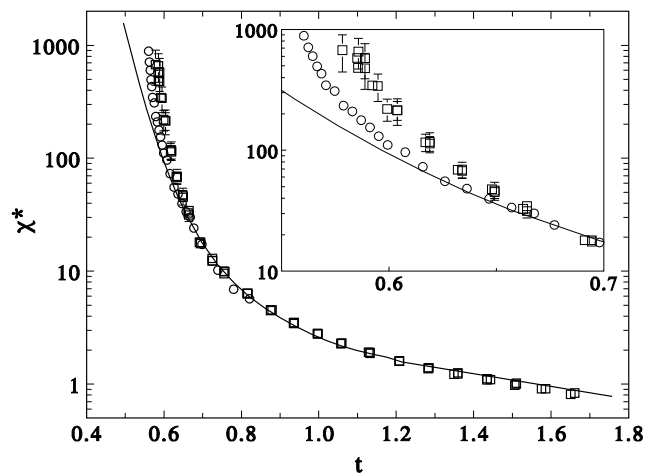,width=82mm,angle=0}}
\caption{
Staggered susceptibility $\chi^*\equiv \chi/\widetilde{S}^2$ 
vs $t$ for $S=5/2$. Symbols and inset as in Fig.~\protect\ref{f.xis25}.
\label{f.chis25}
}
\end{figure}

Figs.~\ref{f.xichis10}-\ref{f.chis25} 
show our results together with the available 
QMC and experimental data. 
We underline that no best-fit procedure is involved in such comparison 
as the PQSCHA has no free parameters once $J$ and $S$ are given.

In the $S=1$ case (Fig.~\ref{f.xichis10}) we compare our curves for $\xi$
and $\chi$ with the new QMC data obtained by Harada et al.\cite{HaradaTK98};
such data, which unfortunately do no extend to low temperatures,
do in fact sit on our curves. Also the experimental data
for La$_2$NiO$_4$ and K$_2$NiF$_4$,
which are not included in Fig.~\ref{f.xichis10} for the sake of clarity,
very well agree with our PQSCHA curves as shown in 
Ref.~\onlinecite{CTVV96prl}.

For $S=5/2$ we use the experimental data relative to 
KFeF$_4$\cite{FultonEtal94} and also those for Rb$_2$MnF$_4$ very recently
obtained by Lee et al.\cite{LeeEtal98}.
Both compounds are characterized by an Ising-like 
anisotropy $H^{\rm A}\sim J/20$: 
Following the reasoning by Birgeneau\cite{Birgeneau90},
the crossover between Ising and Heisenberg behaviour should occur 
when $h^{\rm A}\xi^2\sim 1$, with $h^{\rm A}\equiv H^{\rm A}/(zJS)\sim 0.005$.
These materials are hence expected to behave like 2DQHAF for $\xi\lesssim 14$,
i.e. $t\gtrsim 0.6$, which is in fact the region where the PQSCHA curves 
agree with the experimental data, as seen in 
Figs.~\ref{f.xis25}-\ref{f.chis25}.

As for the compound KFeF$_4$ a few more comments are in order; 
its magnetic ions are distributed on a non perfect square lattice
and the magnetic interaction is hence characterized by two different
exchange integrals $J_a$ and $J_b$; such difference is small
enough to allow the system to be described by Eq.~(\ref{e.H}) but
with the value of $J$, as from different experiments, slightly variable. 
In particular from susceptibility, Raman scattering, and neutron scattering
measurements $J$ is found to be $2.30$~meV, $2.35$~meV, and $2.44$~meV, 
respectively. Neutron scattering experiments by Fulton et
al.\cite{FultonEtal94}
measured $J_a$ and $J_b$ separately getting $J_a=2.18$~meV 
and $J_b=2.73$~meV at $50$~K and $J_a=J_b=2.40$~meV at
$100$~K. In our theory, on the other hand, there is no free parameter and 
the exchange integral $J$ only enters the definition of the reduced
temperature; its value can be hence easily derived by optimizing 
the agreement between the experimental data for the correlation length 
and our curve. In this procedure, in order to take into account the 
above mentioned effects of the Ising anisotropy, we have only used the
experimental data with $\xi<14$, obtaining the value $J=2.33\pm 0.03$~meV;
this result does not change discarding the data with $8<\xi<14$ and agrees 
with the above mentioned independent determinations.

From the extensive comparison between our curves and all the available 
experimental and QMC data, as well as with the HTE results as shown in 
Ref.~\onlinecite{CTVV96prl}, it is evident that the PQSCHA leads to a proper 
description of the thermodynamic behaviour of the QHAF. 

Let us now consider the temperature dependence of the correlation length in
the $S=1$ case (a similar analysis can be developed for any $S\ge 1$):
Our data extend up to $\xi\simeq 10^5$ , i.e. $t\simeq 0.2$, where 
the renormalization coefficient ${\cal{D}}(S=1)$ is still less than $0.45$.
This means that we can safely assume our curves to reproduce the correct
$\xi$ for any $t\ge 0.2$. 
Notice that this lower limit is set by the absence of classical values
for $\xi>10^5$. 

Having said that, we try to fit our curves with the low-temperature 
CHN-HN formula
\begin{equation}
\xi={e\over 8}\left({c\over2\pi\rho}\right)
 \exp \left( 2\pi\rho\over T\right)~
 \left[1-{T\over4\pi\rho}\right]
\label{e.xiCH2N2}
\end{equation}
where $\rho$ and $c$ are the two fitting parameters that do not depend upon
$T$.
In order to assert that Eq.~(\ref{e.xiCH2N2}) describes the correct
temperature dependence of $\xi$ in the temperature range of interest,
the fit must be stable, in the sense that the resulting values of $\rho$
and $c$ must not vary if the fit is restricted to lower or higher
temperatures. In what follows we show that, in fact, this is not the case.

\begin{figure}[hbt]
\centerline{\psfig{bbllx=16mm,bblly=69mm,bburx=192mm,bbury=207mm,%
figure=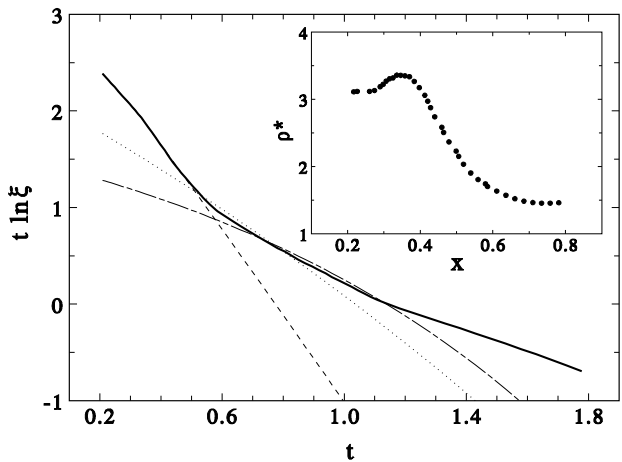,width=82mm,angle=0}}
\caption{$t\ln\xi$ vs $t$ for $S=1$: fits of the PQSCHA curve (full line) 
with Eq.~(\protect\ref{e.xiCH2N2}) in the temperature interval $[x,1.7]$
with $x=0.2$ (dashed line), $0.46$ (dotted line) and $0.56$
(dash-dotted line). The inset shows $\rho\equiv 2\pi\rho/\widetilde{S}^2$ 
as from fits in the temperature interval $[x,1.7]$ 
as a function of $x$.
\label{f.fit}
}
\end{figure}

We have fitted our curve with Eq.~(\ref{e.xiCH2N2}) in different intervals
of temperature and found that the values of $\rho$ and $c$ change 
drastically when the low temperature data are not included in the fit.
To enlighten the discrepancies in Fig.~\ref{f.fit} we report
the quantity $t\ln \xi$ as a function of $t$; the full line is the PQSCHA
curve while the other curves are given by Eq.~(\ref{e.xiCH2N2}) with $\rho$
and $c$ as from the fit in the five temperature intervals 
$[x,1.7]$ with $x=0.2$, $0.46$ and $0.56$. The inset
shows $\rho^*\equiv 2\pi\rho/\widetilde{S}^2$, as resulting from 
fits in the temperature intervals $[x,1.7]$, as a function of $x$.
The fit is clearly unstable and Eq.~(\ref{e.xiCH2N2}) cannot reproduce
the correct behaviour of the 2DQHAF correlation length in the whole 
temperature range.

From this analysis we conclude that the CHN-HN theory does not 
describe the QHAF when $\xi\le 10^5$, i.e. for $t\ge 0.2$ for 
$S=1$ and, more in general, for $t\ge t_0$ where $t_0$ is such that
$\xi_{\rm cl}(t_0/\theta(t_0))\simeq 10^5$ for any $S\ge 1$;
the only reason why Eq.~(\ref{e.xiCH2N2}) seems to reproduce 
the experimental data is that the temperature range where the fit 
is carried out is small enough not to show significant deviations.

The  CHN-HN theory is in fact already known to describe the QHAF only
at sufficiently low temperatures, but no quantitative indication is given
about the actual range of validity, save the fact that this is narrower for
larger spin. What our work clearly shows is that such range of validity,
for $S\ge 1$, does not overlap with the region where experimental 
and QMC data are available; in other terms, such data are not described by 
the three-loop renormalization-group expressions relative to 
the QNL$\sigma$M.

Although our analysis cannot be extended to all temperatures in 
the $S=1/2$ case, we point out that conclusions similar to ours have 
been recently drawn by Beard et al.\cite{BeardEtal97} from their 
low-temperature QMC results. 
On the other hand, Kim and Troyer,
who also performed QMC simulations on the 2DQHAF with $S=1/2$,
claim\cite{KimT98} that their results are in "excellent agreement" with 
the QNL$\sigma$M field-theory predictions.
Nevertheless, such agreement involves unstable fitting
procedures; in particular, the uniform susceptibility $\chi_{\rm u}$
can be fitted to the CHN-HN expression only for $t\le 0.23$ while
for the correlation length the restriction $\xi\ge 39.2$ (i.e. 
$t\le 0.27$) is necessary to make the fit stable but not even 
sufficient to let the resulting values of $\rho$ and $c$
coincide with those obtained by fitting $\chi_{\rm u}$.
In fact we think that the results presented by Kim and Troyer, 
which moreover extend to temperatures not as low as those studied by 
Beard et al., do not
"confirm the validity of the mapping from the QHAF to the NL$\sigma$M",
but rather suggest that also for $S=1/2$ the CHN-HN formulas do not properly
describe the behaviour of the 2DQHAF for $t\gtrsim 0.25$, 
i.e., in the temperature region where most experimental and QMC
data are available.
\medskip

We are grateful to Prof. R.J. Birgeneau and to Y.S. Lee (MIT) for
fruitful exchanges and for providing us with experimental data
prior to publication.


\end{document}